\documentclass[12pt]{article}

\usepackage{amsmath}
\usepackage{amsfonts}
\usepackage{amssymb}
\usepackage{graphicx}
\usepackage{psfrag}
\usepackage{graphicx}
\usepackage{psfrag}

\newcommand{\be}{\begin{equation}}
\newcommand{\ee}{\end{equation}}
\newcommand{\ba}{\begin{eqnarray}}
\newcommand{\ea}{\end{eqnarray}}

\begin{document}
\begin{center}
{\bf
EQUIDISTANCE OF THE COMPLEX 2-DIM ANHARMONIC OSCILLATOR SPECTRUM: EXACT SOLUTION}\\
\vspace{1cm}
{\large \bf F. Cannata $^{1,}$\footnote{E-mail: cannata@bo.infn.it},
M. V. Iof\/fe $^{2,}$\footnote{E-mail: m.ioffe@pobox.spbu.ru},
D. N. Nishnianidze} $^{2,3,}$\footnote{E-mail: cutaisi@yahoo.com}\\
\vspace{0.5cm}
$^1$ INFN, Via Irnerio 46, 40126 Bologna, Italy.\\
$^2$ Saint-Petersburg State University,198504 Sankt-Petersburg, Russia\\
$^3$ Akaki Tsereteli State University, 4600 Kutaisi, Georgia
\end{center}
\vspace{0.5cm}
\hspace*{0.5in}
\vspace{1cm}
\hspace*{0.5in}
\begin{minipage}{5.0in}
{\small
We study a class of quantum two-dimensional models with complex potentials of specific form. They can be considered as the generalization of
a recently studied model with quadratic interaction not amenable to conventional separation of variables. In the present case, the property
of shape invariance provides the equidistant form of the spectrum and the algorithm to construct eigenfunctions analytically. It is shown that the
Hamiltonian is non-diagonalizable, and the resolution of identity must include also the corresponding associated functions.
In the specific case of anharmonic second-plus-fourth order interaction, expressions for the wave functions and associated functions are constructed explicitly for the lowest levels, and the recursive algorithm to produce higher level wave functions is given.
}
\end{minipage}


\section*{\normalsize\bf 1. \quad Introduction.}

Considerable attention has been paid recently to the quantum models with complex potentials, and in particular, to the models
with real spectra \cite{bender0}. Although analysis of such systems is fraught with many difficulties
\cite{bender-review}, \cite{mostafa-review}, it seems to be very useful for modern development both of
generalized Quantum Mechanics and Quantum Field Theory \cite{bender-qft}. In particular, a modified scalar product which provides unitary evolution
must be built, and diagonalization of the Hamiltonians has to be controlled \cite{non-diag}. Very interesting results were obtained
for a wide class of models with unbroken $PT-$invariance \cite{bender0}, \cite{bender01} - \cite{tateo}.
The important notion of the pseudo-Hermiticity was introduced \cite{most} - \cite{japaridze}:
\be
\eta H \eta^{-1} = H^{\dagger} \label{ps}
\ee
with $\eta$ a Hermitian invertible operator. It allowed to define a class
of non-Hermitian systems with physically acceptable properties of energy spectra.
Systematically pseudo-Hermiticity has been presented by
A.Mostafazadeh \cite{most} (see also \cite{ahmed} - \cite{geyer}).
The Hilbert space for such kind of systems is described in terms of a
biorthogonal basis, which consists of the eigenstates $|\Psi_n\rangle$ and $|\tilde\Psi_n\rangle$
of $H$ and $H^{\dagger},$ correspondingly. The essential peculiarity of non-Hermitian models is known:
some of them are naturally described by non-diagonalizable Hamiltonians \cite{non-diag}. For such systems, the usual biorthogonal basis
does not provide a complete basis in Hilbert space, and one has to
add the so-called associated functions to complete the basis. Then, the Hamiltonian is block-diagonal with
some number of Jordan blocks of standard structure on its diagonal.

It must be noticed, that during last years almost all papers in this line were limited to one-dimensional Quantum Mechanics, where the problems have more chance to be fully solved. Nevertheless, some investigations of two-dimensional systems were performed as well
\cite{bender2} - \cite{pseudo}. Recently, a new non-trivial two-dimensional model with complex potential was studied in detail in paper \cite{CIN-osc}. The potential of this model had the simple form of second order polynomial
in $x_1,x_2.$ Usually models of such kind are solved by means of linear transformation of coordinates
with subsequent separation of variables (see for example, \cite{srilanka}), but for two values of coupling constants,
the Hamiltonians are not amenable to
separation of variables. Just such model was studied in \cite{CIN-osc}. Exact solution was possible essentially due to shape invariance property
\cite{shape} - \cite{exact} originated from the Supersymmetrical Quantum Mechanics approach \cite{witten}, \cite{cooper}.
For the model \cite{CIN-osc}, the energy spectrum was proven to be equidistant, the same as for usual real harmonic oscillator. The corresponding wave functions were found analytically as well. From the properties of wave functions it became clear that the Hamiltonian is non-diagonalizable, and its structure was
analyzed in detail. The corresponding associated functions were also built analytically and their properties were studied.

In the present paper, the generalization of the model \cite{CIN-osc} is given: new exactly solvable models are built with the full spectrum and the wave functions analytically calculated. In Section 2, we consider the wide class of complex two-dimensional potentials which include two terms: an arbitrary polynomial function of $\bar z=x_1-ix_2$ in addition to real two-dimensional isotropic harmonic oscillator. Somehow unexpectedly many properties of this system will turn out to be analogous to those of complex quadratic potential of \cite{CIN-osc}. Namely,
the equidistance of spectrum and analytical expressions for wave functions follow directly from the shape invariance. In this sense, the situation is very different from that of the Hermitian case, where the spectrum of anharmonic oscillator (which does not obey any shape invariance property) can be very far from the equidistant. In Section 3, the different properties of the system are studied, such as its pseudo-Hermiticity and modified scalar product. The self-orthogonality of all excited wave functions signals non-diagonalizability of the Hamiltonian with all related features. The analysis leads to the specific algorithm for construction of associated functions for all Jordan blocks for the particular case of specific quartic anharmonic form of interaction. In Section 4, two lowest Jordan cells are studied explicitly, all associated functions are constructed, and scalar products are shown to satisfy the standard rules for non-diagonalizable pseudo-Hermitian Hamiltonians.

\section*{\normalsize\bf 2.\quad Description of the model.}

Let us start from the two-dimensional Hamiltonian with complex interaction of the form:
\ba
H&=&-\Delta^{(2)}+V(\vec x)=-4\partial_z\partial_{\bar z}+\lambda^2z\bar z + 2\lambda\bar zF^{\prime}(\bar z);
\label{H}\\
z&=&x_1+ix_2;\,\,\bar z=x_1-ix_2;
\quad F^{\prime}(\bar z)=\partial_{\bar z}F(\bar z)\nonumber
\ea
where $F(\bar z)$ is supposed to be a polynomial.
This potential is the direct generalization of interaction in \cite{CIN-osc}, where $F^{\prime}$ was proportional
to $\bar z$ with a coefficient such that the conventional separation of variables is impossible. Also the present case (\ref{H}) is not amenable to separation of variables \cite{miller}, \cite{eisenhart}, and the spectrum of the model again can be found by means of shape invariance \cite{shape}, \cite{new} - \cite{shape-our}. This property was introduced \cite{shape} in the framework of Supersymmetrical
Quantum Mechanics as
\be
H^{(2)}(\vec x; \tilde a) = H^{(1)}(\vec x; a)-\mathcal{R}(a) \label{int2}
\ee
between two partner Hamiltonians which are intertwined by supercharge operators $Q^{\pm}:$
\be
H^{(1)}(x; a)Q^{+}(x)=Q^{+}(x)H^{(2)}(x; a); \qquad Q^{-}(x)H^{(1)}(x; a)=H^{(2)}(x; a)Q^{-}(x). \label{int1}
\ee
Above, $a$ and $\tilde a=\tilde a(a)$ are some parameters which define coupling constants of $H^{(1)},\,H^{(2)},$ and $\mathcal R(a)$ does not depend on $x.$ Shape invariance allows to find algebraically the spectrum and wave functions of Hamiltonians - completely in one-dimensional case \cite{shape}, \cite{mallow}, and partially - in two-dimensional situation \cite{new}, \cite{ioffe1}, \cite{shape-2011}.

The simplest realization of one-dimensional shape invariance with $\tilde a=a,$ i.e. with $H^{(2)}(x, a)=H^{(1)}(\vec x; a) + 2 \lambda ,$ was considered in \cite{shape-our}. For that case, both systems were shown to have oscillator-like (equidistant) spectra. The same kind of
shape invariance and of spectra appeared in two-dimensional systems with complex quadratic interaction \cite{CIN-osc}. The basic origin for this property of spectra are the following - oscillator-like - commutation relations for $H :$
\be
HA^+=A^+(H+2\lambda );\quad HA^-=A^-(H-2\lambda ).\label{commutator2}
\ee
These relations are the particular case of (\ref{int1}), where $A^{\pm}$ play the role of supercharges $Q^{\pm}$, $2\lambda$ - the role of $\mathcal R(a),$ and $H,\,H + 2\lambda $ - the role of partner Hamiltonians $H^{(1),(2)}.$
In general, an arbitrary system with Hamiltonian obeying commutation relations (\ref{commutator2}) has oscillator-like equidistant spectrum \cite{shape-our}, maybe on a finite interval of energy: its lowest level is defined by the zero mode of $A^-$ and the highest level - by the zero mode of the operator $A^{+}.$ Since the intertwined Hamiltonians in (\ref{commutator2}) coincide up to a shift,
they are called self-isospectral. Thus, both for general one-dimensional \cite{shape-our} and for specific two-dimensional \cite{CIN-osc} cases, self-isospectrality leads to an equidistant (oscillator-like) character of the spectrum.

Let us consider now in detail the system with Hamiltonian (\ref{H}) . Due to absence of singularities and similarly to the model \cite{CIN-osc},
no "fall to the center" phenomena \cite{landau} is possible here (see the formal proof in Appendix of \cite{CIN-osc}).
The normalizable bound state wave functions will be exponentially
decreasing at infinity, having no singularities. The corresponding spectrum is bounded from below, and the ground state
with energy $E_0$ will be denoted as $\Psi_{0,0}(\vec x)$ (it will be clear below, why we
use two indices for enumeration of $\Psi $).

First of all, the Hamiltonian (\ref{H}) satisfies the commutation relations (\ref{commutator2}) with first order operators $A^{\pm}$ of the form:
\be
A^{\pm}=\partial_z\mp\frac{\lambda}{2}\bar z .
\label{A}
\ee
Similarly to \cite{shape-our},\cite{CIN-osc}, it is necessary to define the zero modes of the intertwining operator $A^-:$
\begin{equation}\label{zero}
A^-\Psi_{n,0}(z, \bar z)=0.
\end{equation}
Looking for the zero modes in the form:
\begin{equation}\label{zero2}
\Psi_{n,0}(z, \bar z)=\exp{\biggl(-\frac{\lambda}{2}z\bar z - F(\bar z)\biggr)}\widetilde\Psi_{n, 0},
\end{equation}
one easily obtains that $\widetilde\Psi_{n, 0}$ depend only on $\bar z.$ The zero mode $\Psi_{n, 0}(z, \bar z)$ to be simultaneously eigenfunction of $H$ with energy $E_n$ has to satisfy the simple equation for $\widetilde\Psi_{n, 0}(\bar z)$:
\begin{equation}
2\lambda\biggl[\bar z\partial_{\bar z}+1\biggr]\widetilde\Psi_{n, 0}(\bar z)=E_n\widetilde\Psi_{n, 0}(\bar z). \nonumber
\end{equation}
Its solutions are:
\begin{equation}\label{unique}
\widetilde\Psi_{n, 0}(\bar z)=c_{n,0}\bar{z}^{\frac{E_n-2\lambda}{2\lambda}},
\end{equation}
where $c_{n, 0}$ are constants. Since the wave functions must be the single-valued functions on a plane, i.e. to be $2\pi-$periodic in polar
angle $\varphi,$ the allowed energy spectrum of the model is:
\begin{equation}\label{spectrum}
E_n=2\lambda (n+1);\quad n=0, 1, 2, ... .
\end{equation}
The ground state corresponds to $n=0,$ and excited states - to $n=1,2,... .$ One can argue that no additional bound states are possible besides (\ref{spectrum}). Indeed, supposing that such additional level with energy $E$ exists, one may act several times on its wave function by the operators $A^-.$ According to the second relation in (\ref{commutator2}), one will obtain wave functions with energies $E-2\lambda m;\, m=1,2,... .$ This series can be cut only if the zero mode of $A^-$ appears on some stage, but all zero modes are described by (\ref{zero2}), (\ref{unique}) above. Therefore, the ground state of the model is:
\begin{equation}\label{zero3}
\Psi_{0,0}(z, \bar z)=c_{0, 0}\exp{\biggl(-\frac{\lambda}{2}z\bar z - F(\bar z)\biggr)};\quad E_0=2\lambda .
\end{equation}
One can check by direct calculation that wave functions of excited states given by (\ref{zero2}), (\ref{unique}) can be also written as:
\begin{equation}\label{nlevel}
\Psi_{n,0}(z, \bar z)=c_{n,0}\bar z^n\exp{\biggl(-\frac{\lambda}{2}z\bar z - F(\bar z)\biggr)} = (A^+)^n\Psi_{0,0}(z, \bar z),\,\, E_n=2\lambda (n+1);\, n=1,2,... ,
\end{equation}
with $c_{n, 0}\equiv (-\lambda /2)^nc_{0,0}.$

Thus, the whole spectrum (\ref{spectrum}) and the corresponding wave functions $\Psi_{n,0}(z,\bar z)$ of the system (\ref{H}) are known
analytically. It should be noted that the spectrum (\ref{spectrum}) depends neither on the structure of function $F(\bar z)$ nor on its coupling constants (coefficients $b$ and $\omega$ below Eqs.(\ref{27}) and (\ref{31})). This is a general consequence of relations (\ref{commutator2}).  A further investigation of properties of the wave functions will be performed in the next Section.

\section*{\normalsize\bf 3.\quad Non-diagonalizability.}

It is known \cite{bender-review}, \cite{mostafa-review} (see also \cite{CIN-osc}) that the self-consistent formulation of Quantum Mechanics with pseudo-Hermitian (\ref{ps}) Hamiltonians includes a suitable modification of the scalar product and resolution of identity. Namely,
a new scalar product has to be used:
\begin{equation}\label{B}
\langle\langle\Psi |\Phi\rangle\rangle \equiv \int (\eta T\Psi)\Phi ,
\end{equation}
so that the pseudo-Hermitian $H$ becomes Hermitian when equipped with (\ref{B}). In our present case (\ref{H}), one may define $\eta \equiv P_2,$ the reflection $x_2\leftrightarrow -x_2,$ and therefore, the scalar product (\ref{B}) is simply an integral over the
product $\int\Psi \Phi ,$ instead of the $\int\Psi^{\star}\Phi$ in the ordinary Quantum Mechanics, where $\eta \equiv 1.$

First of all, we have to check the properties of all wave functions (\ref{nlevel}) in the framework of
a new scalar product. Since below we will need the results of integration of some functions over $z, \bar z,$ the list of such integrals
is given in Appendix. In particular, the norm of states $\Psi_{n, 0}$ vanishes for $n\geq 1:$
\ba
&&\langle\langle\Psi_{n,0} |\Psi_{n,0}\rangle\rangle = \int (\Psi_n)^2d^2x =
\frac{c^2_{n,0}}{2}\int \bar z^{2n}\exp{[-\lambda z\bar z -2 F(\bar z)]}dzd\bar z = \nonumber\\
&&=\frac{\pi c^2_{n,0}}{\lambda }\exp{(-2F(0))}\delta_{n0}
\label{norm}
\ea
due to expansion of $\exp{(2F(\bar z))}$ in powers of $\bar z$ (see the list of integrals in Appendix).

Thus, the usual resolution of identity in terms of the so-called biorthogonal basis \cite{mostafa-review}, \cite{most}, \cite{levai}, \cite{pseudo}, \cite{CIN-osc} $|\Psi_n\rangle,\,|\tilde\Psi_n\rangle$ does not hold, and the Hamiltonian is non-diagonalizable. Details can be found in \cite{non-diag} in a one-dimensional case, and in \cite{CIN-osc} in a two-dimensional context: it is necessary
to build the so-called associated functions which participate to the resolution of identity and complete the basis.

For such systems,
each self-orthogonal
wave function $\Psi_{n,0},\, n\geq 1$ with zero norm must be accompanied with a set of $p_n-1$
associated functions $\Psi_{n,k},\, k=1,2,...,p_n-1,$ where $p_n$ is called the dimension of Jordan cell. This situation (c.f. Eq.(\ref{assoc}) below) is very different from the familiar one with degeneracy of energy level of Hermitian Hamiltonian. It should be clear now, why
notations with two indices of wave functions were introduced above. By definition, these functions obey:
\begin{equation}\label{assoc}
 (H-E_n)\Psi_{n,k}=\Psi_{n,k-1};\quad k=1,2,...,p_n-1,
\end{equation}
where all functions are supposed to be normalizable.
Each self-orthogonal eigenfunction $\Psi_{n,0},\, n=1,2,...$
is supposed to be accompanied by {\it only one} set of associated functions $\Psi_{n,k},\, k=1,2,...,p_n-1.$

Similarly to the scheme of the previous paragraph, the partner eigenfunctions
$\tilde\Psi_{n,0}$ also are accompanied by their associated functions $\tilde\Psi_{n,k},\, k=1,2,...,p_n-1.$
The following numeration for the functions $\tilde\Psi $ is convenient:
\begin{equation}\label{renumber}
\tilde\Psi_{n,p_n-k-1}=\Psi^{\star}_{n,k}\quad k=0,1,2,...p_n-1.
\end{equation}
With these notations, according to the general formalism which was illustrated in detail for some one-dimensional
models \cite{non-diag}, the scalar product in the extended biorthogonal basis is:
\ba
&&\langle\langle\Psi_{n,k}|\Psi_{m,l}\rangle\rangle = \langle\tilde\Psi_{n,k}|\Psi_{m,l}\rangle =
\int \Psi_{n,k}(\vec x)\Psi_{m,l}(\vec x)d^2x =\delta_{nm}\delta_{k\, (p_n-l-1)};\label{product}\\
&&k=0,1,...,p_n-1;\,\, l=0,1,...,p_m-1 .\nonumber
\ea
Correspondingly, the generalized decompositions become:
\ba
I&=&\sum_{n=0}^{\infty}\sum_{k=0}^{p_n-1} |\Psi_{n,k}\rangle\rangle\langle\langle\Psi_{n,p_n-k-1}|;\label{decomp1} \\
H&=&\sum_{n=0}^{\infty}\sum_{k=0}^{p_n-1} E_n|\Psi_{n,k}\rangle\rangle\langle\langle\Psi_{n,p_n-k-1}|
+ \sum_{n=0}^{\infty}\sum_{k=0}^{p_n-2} |\Psi_{n,k}\rangle\rangle\langle\langle\Psi_{n,p_n-k-2}|.\label{decomp2}
\ea
The Hamiltonian $H$ is clearly non-diagonal, but block-diagonal. Each block - Jordan cell of standard form
(see (\ref{decomp2})) - has dimensionality $p_n,$ and we assume the absence of degeneracy: each eigenvalue corresponds to
one Jordan cell.

The crucial property of eigenfunctions $\Psi_{n,0}$ - their self-orthogonality for $n \geq 1$ - was already displayded in (\ref{norm}).
To complete the construction of the Jordan cell, it is necessary to find the corresponding associated functions with properties listed above.

The scalar products (\ref{product}) vanish for different energy levels $E_n,\,E_m,$ i.e.
$\langle\langle\Psi_{n,k}|\Psi_{m,l}\rangle\rangle =0$ for $n\neq m.$
For the proof, the pseudo-Hermiticity of $H$ is important. For example, similarly to the  ordinary Quantum Mechanics,
wave functions with different energies are orthogonal:
$$
0=\langle\langle\Psi_{n,0}|H|\Psi_{m,0}\rangle\rangle-\langle\langle H\Psi_{n,0}|\Psi_{m,0}\rangle\rangle
=(E_m-E_n)\langle\langle\Psi_{n,0}|\Psi_{m,0}\rangle\rangle .
$$
Further, because of (\ref{assoc}):
\ba
0&=&\langle\langle\Psi_{n,0}|(H-E_m)|\Psi_{m,1}\rangle\rangle -
\langle\langle (H-E_m)\Psi_{n,0}|\Psi_{m,1}\rangle\rangle=\nonumber\\&=&
\langle\langle \Psi_{n,0}|\Psi_{m,0}\rangle\rangle -(E_n-E_m)\langle\langle\Psi_{n,0}|\Psi_{m,1}\rangle\rangle \nonumber ,
\ea
and the scalar products between wave functions and first associated functions for different $E_n,\,E_m$ also vanish:
$$
\langle\langle \Psi_{n,0}|\Psi_{m,1}\rangle\rangle = \langle\langle\Psi_{n,1}|\Psi_{m,0}\rangle\rangle =0.
$$
The procedure can be continued leading to orthogonality of all functions with different $n,m.$

Such indirect method does not allow to study the scalar products of associated functions with the same value $n:$ it is necessary to
build them explicitly. This task is more difficult than in the case of harmonic oscillator \cite{CIN-osc}.
Let us transform the defining equations to the form which can provide some simple algorithm to construct associated functions.
It follows directly from the commutation relations (\ref{commutator2}), that
\begin{equation}
(A^-)^k(H-E_n)=(H-E_{n-k})(A^-)^k.   \nonumber
\end{equation}
Together with relation (\ref{assoc}), for $k=1$ it gives:
\be
(H-E_{n-1})A^-\Psi_{n,1}=0, \nonumber
\ee
and therefore, the wave function and its first associated are related:
\be
A^-\Psi_{n,1}=a_{n,1}\Psi_{n-1, 0};\,\, a_{n,1}=Const.
\label{nn-1}
\ee
Iterating this procedure, one obtains:
\be
(A^-)^k\Psi_{n,k}=a_{n,k}\Psi_{n-k, 0};\,\, a_{n,k}=Const,
\label{nk}
\ee
with the last relation of this sort:
\be
(A^-)^n\Psi_{n,n}=a_{n,n}\Psi_{0, 0}.
\label{nn}
\ee
Functions in r.h.s. of (\ref{nn-1}) - (\ref{nn}) are known from (\ref{nlevel}), and the ansatz similar to (\ref{zero2})
\be
\Psi_{n,k}=\exp{\bigl(-\frac{\lambda}{2}z\bar z-F(\bar z)\bigr)}\widetilde\Psi_{n,k} \nonumber
\ee
is useful to look for all associated functions. Indeed, one can check that
\be
(A^-)^m\exp{\bigl(-\frac{\lambda}{2}z\bar z-F(\bar z)\bigr)}=\exp{\bigl(-\frac{\lambda}{2}z\bar z-F(\bar z)\bigr)}\partial_z^m,
\nonumber
\ee
providing the following inhomogeneous differential equation:
\be
\partial_z^k\widetilde\Psi_{n,k}=a_{n,k}c_{n-k,0}\bar z^{n-k}.
\nonumber
\ee
Its general solution is the polynomial in $z:$
\be
\widetilde\Psi_{n,k}=\frac{a_{n,k}c_{n-k,0}}{k!}z^k\bar z^{n-k} +\sum_{i=0}^{k-1}g_{n,k}^{(i)}(\bar z)z^i,
\label{general}
\ee
but the functions $g_{n,k}^{(i)}(\bar z)$ are yet arbitrary functions of $\bar z.$ These functions have to be found by substitution of (\ref{general}) into relations (\ref{assoc}) connecting $\Psi_{n,k}$ with $\Psi_{n,k-1}$. Due to:
\be
(H-E_n)\exp(-\frac{\lambda}{2}z\bar{z}-F(\bar{z}))=2\exp(-\frac{\lambda}{2}z\bar{z}-F(\bar{z}))
\biggl(-2\partial_z\partial_{\bar{z}}+\lambda\bar{z}\partial_{\bar{z}}+\lambda z\partial_z+
2F'(\bar{z})\partial_z-\lambda n\biggr),\nonumber
\ee
the equations for $g_{n,k}^{(i)}(\bar z)$ take the form:
\ba
& &2\biggl(-2\partial_z\partial_{\bar{z}}+\lambda\bar{z}\partial_{\bar{z}}+\lambda z\partial_z+
2F'(\bar{z})\partial_z-\lambda n\biggr)\biggl(\frac{a_{n,k}c_{n-k,0}}{k!}z^k\bar{z}^{n-k}+
\sum_{i=0}^{k-1}g^{(i)}_{n,k}(\bar{z})z^i\biggr)=\nonumber\\
& &=\biggl(\frac{a_{n,k-1}c_{n-k+1,0}}{(k-1)!}z^{k-1}\bar{z}^{n-k+1}+
\sum_{i=0}^{k-2}g^{(i)}_{n,k-1}(\bar{z})z^i\biggr).\label{22}
\ea
Considering (\ref{22}) as polynomial in $z,$ one has trivial identity in highest power $z^k,$ but for the next powers, (\ref{22}) gives the system of coupled first order differential equations for unknown function $g_{n,k}^{(i)}(\bar z),$ which are written separately for $i=k-1$ as:
\be
2\lambda\biggl[\bar{z}^{k-n-1}g^{(k-1)}_{n,k}(\bar{z})\biggr]^{\prime}=\frac{a_{n,k-1}c_{n-k+1,0}}{(k-1)!}\bar{z}^{-1}+
\frac{4a_{n,k}c_{n-k,0}}{(k-1)!}\biggl[(n-k)\bar{z}^{-3}-F'(\bar{z})\bar{z}^{-2}\biggr], \label{25}
\ee
and for $i=0,1,...,(k-2)$ as:
\be
2\lambda\biggl[\bar z^{i-n}g^{(i)}_{n,k}(\bar z)\biggr]^{\prime}=\biggl[4(i+1)\biggl((g^{(i+1)}_{n,k})^{\prime}(\bar z)-F^{\prime}(\bar z)
g^{(i+1)}_{n,k}(\bar z)\biggr)+g^{(i)}_{n,k-1}(\bar{z})\biggr]\bar{z}^{i-n-1}. \label{26}
\ee

It is convenient to extract explicitly the quadratic term from the function $F(\bar z):$
\be
F(\bar z)=\frac{b}{2}\bar z^2+f(\bar z).
\label{27}
\ee
Then, equation (\ref{25}) takes the form:
\be
2\lambda\biggl[\bar z^{k-n-1}g^{(k-1)}_{n,k}(\bar z)\biggr]^{\prime}=\frac{a_{n,k-1}c_{n-k+1, 0}-4ba_{n,k}c_{n-k,0}}{(k-1)!}\bar z^{-1}+
\frac{4a_{n,k}c_{n-k,0}}{(k-1)!}\biggl[(n-k)\bar z^{-3}-f^{\prime}(\bar z)\bar z^{-2}\biggr]
\label{28}
\ee
Since the wave functions, and therefore the functions $g^{(k-1)}_{n,k},$ must be single-valued on a plane, we have to exclude the logarithmic
term from the solution of (\ref{28}). This is possible by choosing the coefficient in the first term in r.h.s. vanishing:
\be
a_{n,k-1}c_{n-k+1, 0}=4ba_{n,k}c_{n-k,0},
\label{29}
\ee
and the solution of (\ref{25}) is:
\ba
g^{(k-1)}_{n,k}(\bar{z})=-\frac{a_{n,k}c_{n-k,0}}{\lambda (k-1)!}\biggl[\alpha^{(k-1)}_{n,k}\bar z^{n-k+1}+(n-k)\bar{z}^{n-k-1}+
2\bar{z}^{n-k+1}\int f^{\prime}(\bar{z})\bar{z}^{-2}d\bar{z}\biggr],\label{30}
\ea
where $\alpha^{(k-1)}_{n,k}$ are integration constants.

Now, (\ref{30}) and (\ref{26}) can be used to find iteratively functions $g^{(i)}_{n,k}(\bar z)$ for $i=k-2,\, i=k-3,$ and so on. Thereby,
the required associated functions $\Psi_{n,k}$ will be obtained. Below, for simplicity we restrict ourselves to the typical quartic anharmonic interaction:
\be
f(\bar z) = \frac{\omega}{2}\bar z^4, \label{31}
\ee
which will clarify the subsequent construction without too cumbersome formulas. With this interaction, expressions (\ref{30}) give the first associated function which depends on the only function $g^{(0)}_{n,1}(\bar z):$
\ba
&\Psi_{n,1}(\bar z)=\biggl[a_{n,1}c_{n-1,0}z\bar z^{n-1}+g^{(0)}_{n,1}(\bar z)\biggr]\exp{\biggl(-\frac{1}{2}(\lambda z\bar z +b\bar z^2+\omega\bar z^4)\biggr)}=\label{A2}\\
&=a_{n,1}c_{n-1,0}\biggl[z\bar z^{n-1}-
\frac{n-1}{\lambda}\bar z^{n-2}-\frac{\alpha^{(0)}_{n,1}}{\lambda}\bar z^n-\frac{2\omega}{\lambda}\bar z^{n+2}\biggr]
\exp{\biggl(-\frac{1}{2}(\lambda z\bar z +b\bar z^2+\omega\bar z^4)\biggr)}.
\nonumber
\ea
Analogously, the second associated function depends on two functions $g:$
\be
\Psi_{n,2}=\biggl[\frac{1}{2}a_{n,2}c_{n-2,0}z^2\bar z^{n-2}+g^{(0)}_{n,2}(\bar z)+g^{(1)}_{n,2}(\bar z)z\biggr]
\exp{\biggl(-\frac{1}{2}(\lambda z\bar z +b\bar z^2+\omega\bar z^4)\biggr)}.
\label{A3}
\ee
Here, $g^{(1)}_{n,2}$ is given by (\ref{30}), but $g^{(0)}_{n,2}$ has to be calculated from (\ref{26}) with $i=0,\,k=2$ and taking into account
relation (\ref{29}):
\ba
&\biggl(\bar z^{-n}g^{(0)}_{n,2}\biggr)^{\prime}=-\frac{2a_{n,2}c_{n-2,0}}{\lambda^2}\biggl[\biggl((n-1)\alpha^{(1)}_{n,2}+b\biggr)\bar z^{-3}
+(n-2)(n-3)\bar z^{-5}+\nonumber\\
&+\biggl(6\omega+b(\alpha^{(0)}_{n,1}-\alpha^{(1)}_{n,2})\biggr)\bar z^{-1}-2\omega\alpha^{(1)}_{n,2}\bar z-4\omega^2\bar z^3\biggr].
\label{A6}
\ea
Again, to avoid the logarithmic term in $g^{(0)}_{n,2}$ the relation
\be
6\omega = b(\alpha^{(1)}_{n,2}-\alpha^{(0)}_{n,1})
\label{missed}
\ee
must be fulfilled.

Thus (\ref{A3}) provides the explicit expression for the second associated function:
\ba
&\Psi_{n,2}(\bar z)=a_{n,2}c_{n-2,0}\Biggl\{\frac{1}{2}z^2\bar z^{n-2}+\label{A8}\\
&+\frac{1}{\lambda^2}\biggl[\biggl((n-1)\alpha^{(1)}_{n,2}+b\biggr)\bar z^{n-2}+\frac{1}{2}(n-2)(n-3)\bar z^{n-4}+2\omega\alpha^{(1)}_{n,2}\bar z^{n+2}+2\omega^2\bar z^{n+4}+2\beta_{n,2}\bar z^n\biggr] -\nonumber\\
&\frac{1}{\lambda}z\biggl(\alpha^{(1)}_{n,2}\bar z^{n-1}+(n-2)\bar z^{n-3}+2\omega\bar z^{n+1}\biggr)\Biggr\}
\exp{\biggl(-\frac{1}{2}(\lambda z\bar z +b\bar z^2+\omega\bar z^4)\biggr)}.\nonumber
\ea
Higher associated functions can be calculated straightforwardly with an analogous procedure.

\section*{\normalsize\bf 4.\quad The lowest Jordan cells.}

For $n=1, 2,$ results (\ref{A2}) and (\ref{A8}) obtained in the previous Section
are already sufficient to know the corresponding Jordan cells completely.
For the first Jordan cell with dimensionality $p_n=n+1=2$, the wave function and associated function are:
\ba
& &\Psi_{1,0}=c_{1,0}\bar z \exp{\biggl(-\frac{1}{2}(\lambda z\bar z +b\bar z^2+\omega\bar z^4)\biggr)}\label{A9}\\
& &\Psi_{1,1}=N_{1,1}\biggl[z-\frac{1}{\lambda}(\alpha^{(0)}_{1,1}\bar z +2\omega\bar z^3)\biggr]
\exp{\biggl(-\frac{1}{2}(\lambda z\bar z +b\bar z^2+\omega\bar z^4)\biggr)};\,\, N_{1,1}=Const. \nonumber
\ea
To be sure, it is necessary to check the scalar products inside the Jordan cell. We know already that the wave function $\Psi_{1,0}$ is
self-orthogonal. The scalar product $\langle\langle\Psi_{1,0}|\Psi_{1,1}\rangle\rangle$ is:
\ba
& &\int\Psi_{1,0}\Psi_{1,1}d^2x=\frac{1}{2}\int\Psi_{1,0}\Psi_{1,1}dzd\bar z=
c_{1,0}N_{1,1}\int\exp{\biggl(-(\lambda z\bar z+b\bar z^2+\omega\bar z^4)\biggr)}\,z\bar z\, dzd\bar z=\nonumber\\
& &=c_{1,0}N_{1,1}\exp{\biggl(-\omega\frac{\partial^2}{\partial b^2}\biggr)}\biggl(-\frac{\partial}{\partial\lambda}\biggr)I(\lambda ,b,c)_{|c=0}=
\frac{2\pi}{\lambda^2}c_{1,0}N_{1,1},
\label{A12}
\ea
where we used the fact that derivatives over $b$ and $\lambda$ commute, and that all integrals $I_{N=0, M}$ vanish due to relations in Appendix. Thus, choosing suitable normalization constants:
\be
\langle\langle\Psi_{1,0}|\Psi_{1,1}\rangle\rangle=1,
\label{A10}
\ee
and, what is important, it does not need fixing the constant $\alpha_{1,1}^{(0)}.$

For the last scalar product in the first Jordan cell, we have to calculate:
\ba
& &\langle\langle\Psi_{1,1}|\Psi_{1,1}\rangle\rangle =
\frac{1}{2}N^2_{1,1}
\exp{\biggl(-\omega\frac{\partial^2}{\partial b^2}\biggr)}\int\exp{\biggl(-(\lambda z\bar z+b\bar z^2)\biggr)}\,\biggl(z^2-\frac{2}{\lambda}\alpha_{1,1}^{(0)}z\bar z\biggr)\, dzd\bar z
=\nonumber\\
& &=\frac{1}{2}N_{1,1}^2\exp{\biggl(-\omega\frac{\partial^2}{\partial b^2}\biggr)}\,\biggl(-\frac{\partial}{\partial c}
+\frac{2\alpha_{1,1}^{(0)}}{\lambda}\frac{\partial}{\partial\lambda}\biggr)\,I(\lambda ,b,c)_{|c=0}=-N^2_{1,1}\frac{2\pi}{\lambda^3}(b+\alpha_{1,1}^{(0)}),
\label{A122}
\ea
where again we used the integrals of Appendix. Here, the integration constant $\alpha_{1,1}^{(0)},$ which was not yet fixed, can be chosen as
$\alpha_{1,1}^{(0)}=-b,$ so that the scalar product (\ref{A122}) vanishes. Therefore, the first Jordan cell has dimensionality $p_{1}=2,$
it includes the wave function and associated function:
\ba
& &\Psi_{1,0}=c_{1,0}\bar z \exp{\biggl(-\frac{1}{2}(\lambda z\bar z +b\bar z^2+\omega\bar z^4)\biggr)}\label{A15}\\
& &\Psi_{1,1}=N_{1,1}\frac{1}{\lambda}\biggl(\lambda z + b\bar z -2\omega\bar z^3\biggr)
\exp{\biggl(-\frac{1}{2}(\lambda z\bar z +b\bar z^2+\omega\bar z^4)\biggr)}. \nonumber
\ea
and all scalar products just correspond to (\ref{product}).

The situation with the next Jordan cell $n=2$ with dimensionality $p_2=3$ can be studied analogously.
\ba
& &\Psi_{2,0}=c_{2,0}\bar z^2 \exp{\biggl(-\frac{1}{2}(\lambda z\bar z +b\bar z^2+\omega\bar z^4)\biggr)}\nonumber\\
& &\Psi_{2,1}=N_{2,1}\biggl[z\bar z-\frac{1}{\lambda}(1+\alpha^{(0)}_{2,1}\bar z^2 +2\omega\bar z^4)\biggr]
\exp{\biggl(-\frac{1}{2}(\lambda z\bar z +b\bar z^2+\omega\bar z^4)\biggr)}; \nonumber\\
& &\Psi_{2,2}=N_{2,2}\Biggl\{\frac{1}{2}z^2+\frac{1}{\lambda^2}\biggl[(\alpha^{(1)}_{2,2}+b)+2\omega\alpha^{(1)}_{2,2}\bar z^4 +2\omega^2\bar z^6
+2\beta_{2,2}\bar z^2\biggr]-\frac{z}{\lambda}(\alpha^{(1)}_{2,2}\bar z+2\omega\bar z^3)\Biggr\}\cdot \nonumber\\
& &\cdot\exp{\biggl(-\frac{1}{2}(\lambda z\bar z +b\bar z^2+\omega\bar z^4)\biggr)}, \nonumber
\ea
where
\be
\alpha^{(1)}_{2,2}-\alpha^{(0)}_{2,1}=6\omega / b,
\label{A17}
\ee
according to (\ref{missed}). Three of scalar products can be calculated directly by means of
the same technique as for the first Jordan cell. The result is:
\ba
& &\langle\langle\Psi_{2,0}|\Psi_{2,2}\rangle\rangle = \pi c_{2,0}N_{2,2} / \lambda^3;
\nonumber\\
& &\langle\langle\Psi_{2,1}|\Psi_{2,1}\rangle\rangle = \pi N^2_{2,1} / \lambda^3;
\nonumber\\
& &\langle\langle\Psi_{2,0}|\Psi_{2,1}\rangle\rangle = 0, \nonumber
\ea
in full agreement with (\ref{product}): the normalization constants $N_{2,2}, N_{2,1}$ can be chosen as necessary.

Two other scalar products can be also brought in correspondence with (\ref{product}), if the integration constants are fixed suitably, and the relation (\ref{A17}) is taken into account. The first of them, up to normalization constant:
\ba
& &\langle\langle\Psi_{2,1}|\Psi_{2,2}\rangle\rangle \sim \int\exp{\biggl(-(\lambda z\bar z+b\bar z^2+\omega\bar z^4)\biggr)}\cdot\nonumber\\
& &\cdot\biggl[ \frac{1}{2}(z\bar z)z^2-\frac{1}{2\lambda}z^2-\frac{2\alpha^{(1)}_{2,2}+\alpha^{(0)}_{2,1}}{2\lambda}(z\bar z)^2+
\frac{2\alpha^{(1)}_{2,2}+b}{\lambda^2}z\bar z-\frac{\alpha^{(1)}_{2,2}+b)}{\lambda^3} \biggr]\, dzd\bar z;
\nonumber
\ea
vanishes, if the constants satisfy the relation
$\alpha^{(0)}_{2,1}+\alpha^{(1)}_{2,2}=-2b,$
which together with (\ref{A17}) gives expressions for constants:
\be
\alpha^{(0)}_{2,1}=-b-\frac{3\omega}{b};\quad \alpha^{(1)}_{2,2}=-b+\frac{3\omega}{b}.
\nonumber
\ee
The second scalar product is:
\ba
& &\langle\langle\Psi_{2,2}|\Psi_{2,2}\rangle\rangle \sim \int\exp{\biggl(-(\lambda z\bar z+b\bar z^2+\omega\bar z^4)\biggr)}\cdot\nonumber\\
& &\cdot\biggl[ \frac{1}{4}z^4+\frac{1}{\lambda^4}(\alpha^{(1)}_{2,2}+b)^2+\frac{(\alpha^{(1)}_{2,2})^2}{\lambda^2}(z\bar z)^2+
\frac{\alpha^{(1)}_{2,2}+b}{\lambda^2}z^2+\frac{2\beta_{2,2}}{\lambda^2}(z\bar z)^2-\nonumber\\
& &-\frac{\alpha^{(1)}_{2,2}}{\lambda}(z\bar z)z^2-
 \frac{2(\alpha^{(1)}_{2,2}+b)\alpha^{(1)}_{2,2}}{\lambda^3}z\bar z-\frac{2\omega}{\lambda}(z\bar z)^3 \biggr]\, dzd\bar z .
 \label{A23}
\ea
The condition, that (\ref{A23}) vanishes, provides the expression for the integration constant:
\be
4\beta_{2,2}=-\frac{9\omega^2}{b^2}+12\omega +b^2,
\nonumber
\ee
and finally, the associated function $\Psi_{2,2}$ can be written in a rather compact form:
\be
\Psi_{2,2} \sim \biggl[ \biggl( \lambda z+(b-\frac{3\omega}{b})\bar z-2\omega\bar z^3 \biggr)^2+18\omega(1-\frac{\omega}{b})\bar z^2+\frac{6\omega}{b} \biggr] \exp{\biggl(-\frac{1}{2}(\lambda z\bar z+b\bar z^2+\omega\bar z^4)\biggr)}.
\label{A26}
\ee
Therefore, for the second Jordan cell all scalar products (\ref{product}) are under control. Thus, we have elaborated the algorithm for construction, step by step, of higher Jordan cells with dimensionalities $p_n=n+1$ and energy $E_n=2\lambda (n+1),$ although the explicit expressions for $\Psi_{n,k}$ will become more and more complicated.

\section*{\normalsize\bf Appendix.}

The set of relevant integrals over $z, \bar z$ was obtained in \cite{CIN-osc}, and it'll be given again below for the reader convenience.
The basic integral \cite{prudnikov}, \cite{CIN-osc} is:
\begin{equation}
I(\lambda ,b,c)=\int \exp{[-(\lambda z\bar z+b\bar z^2+cz^2)]}dzd\bar z=2\pi\delta^{-1};\,\,\delta\equiv \sqrt{(\lambda^2-4bc)}, \nonumber
\end{equation}
with $\lambda > (b+c).$
Then, the next required integrals with power pre-exponential integrand are:
\begin{equation}
I_{N,M}\equiv\int z^N\bar z^M \exp{[-(\lambda z\bar z+b\bar z^2)]}dzd\bar z            \nonumber
\end{equation}
with $\lambda > b$ and integer $N,M.$ These integrals vanish for odd values of $(N+M)$
due to antisymmetry under a space reflection $(x_1,x_2) \rightarrow -(x_1,x_2).$
In turn, for even values of $(N+M)$ the integrals can be calculated by suitable differentiations of $I(\lambda ,b,c)=I(\delta):$
\ba
&&I_{2n,2(n+k)}=\int (z\bar z)^{2n}\bar z^{2k} \exp{[-(\lambda z\bar z+b\bar z^2+cz^2)]}dzd\bar z_{|c=0}=\nonumber\\
&&=[(-\partial_{\lambda})^{2n}(-\partial_b)^{k}I(\delta )]_{|c=0}=0;\,\,k>0 \nonumber\\
&&I_{2(n+k),2n}=\int (z\bar z)^{2n}z^{2k} \exp{[-(\lambda z\bar z+b\bar z^2+cz^2)]}dzd\bar z_{|c=0}=\nonumber\\
&&[(-\partial_{\lambda})^{2n}(-\partial_c)^{k}I(\delta )]_{|c=0}=
\pi (-1)^k2^{k}(2k+1)!(2k+1)_{2n}b^ka^{-(2k+2n+1)} ; \nonumber\\
&&I_{2n+1,2(n+k)+1}=\int (z\bar z)^{2n+1}\bar z^{2k} \exp{[-(\lambda z\bar z+b\bar z^2+cz^2)]}dzd\bar z_{|c=0}=\nonumber\\
&&=[(-\partial_{\lambda})^{2n+1}(-\partial_b)^{k}I(\delta )]_{|c=0}=0; \,\, k>0 \nonumber\\
&&I_{2(n+k)+1,2n+1}=\int (z\bar z)^{2n+1} z^{2k} \exp{[-(\lambda z\bar z+b\bar z^2+cz^2)]}dzd\bar z_{|c=0}=\nonumber\\
&&=[(-\partial_a)^{2n+1}(-\partial_c)^{k}I(\delta )]_{|c=0}=
\pi (-1)^k2^{k}(2k+1)!(2k+1)_{2n+1}b^ka^{-(2k+2n+2)}, \nonumber
\ea
with $(L)_P\equiv L(L+1)...(L+P-1).$
In particular, it is clear that $I_{N,M}=0$ for $M>N.$

\section*{\bf Acknowledgments.}

The work was partially supported by INFN and the University of Bologna (M.V.I. and D.N.N.).


\begin{thebibliography}{}

\bibitem{bender0}
Bender C.M. and Boettcher S. {\it Phys.Rev.Lett.} {\bf 80}  5243 (1998).
\bibitem{bender-review}
Bender C.M. {\it Contemp. Phys.} {\bf 46} 277 (2005);\\
Bender C.M. {\it Rep. Prog. Phys.} {\bf 70} 947 (2007).
\bibitem{mostafa-review}
Mostafazadeh A. and Batal A. {\it J. Phys. A: Math. Gen.} {\bf 37} 11645 (2004);\\
Mostafazadeh A. {\it Int. J. Geom. Meth. Mod. Phys.} {\bf 7} 1191 (2010).
\bibitem{bender-qft}
Bender C.M., Braunchina V., Messina E., arXiv 1201.1244.
\bibitem{non-diag}
Mostafazadeh A. {\it J. Math. Phys.} {\bf 43} 6343 (2002);\\
Scolarici G. and Solombrino L. {\it J. Math. Phys.} {\bf 44} 4450 (2003);\\
Samsonov B.F. and Roy P., {\it J. Phys. A: Math. Gen.} {\bf 38} L249 (2005);\\
Sokolov A.V., Andrianov A.A. and Cannata F. {\it J. Phys. A: Math. Gen.} {\bf 39} 10207 (2006);\\
Andrianov A.A., Cannata F. and Sokolov A.V. {\it Nucl. Phys. B} {\bf 773} 107 (2007).
\bibitem{bender01}
Bender C.M., Boettcher S. and Meisinger P.N. {\it J. Math. Phys.} {\bf 40} 2201 (1999).
\bibitem{bender02}
Bender C.M., Brody D.C. and Jones H.F. {\it Phys. Rev. Lett.} {\bf 89} 270401 (2002).
\bibitem{tateo}
Dorey P., Dunning C. and Tateo R. {\it J. Phys. A: Math. Gen.} {\bf 34} L391 (2001);\\
Dorey P., Dunning C. and Tateo R. {\it J. Phys. A: Math. Gen.} {\bf 34} 5679 (2001).
\bibitem{most}
Mostafazadeh A. {\it J. Math. Phys.} {\bf 43} 205 (2002);\\
Mostafazadeh A. {\it J. Math. Phys.} {\bf 43} 2814 (2002);\\
Mostafazadeh A. {\it J. Math. Phys.} {\bf 43} 3944 (2002).
\bibitem{ahmed}
Ahmed Z. {\it Phys. Lett. A} {\bf 294} 287 (2002).
\bibitem{japaridze}
Japaridze G.S. {\it J. Phys. A: Math. Gen.} {\bf 35} 1709 (2002).
\bibitem{geyer}
Scholtz F.G., Geyer H.B., Hahne F.J.W. {\it Annals of Physics} {\bf 213} 74 (1992).
\bibitem{bender2}
Bender C.M., Dunne G.V., Meisinger P.N. and Simsek M. {\it Phys. Lett. A} {\bf 281} 311 (2001).
\bibitem{srilanka}
Nanayakkara A. {\it Phys. Lett. A} {\bf 304} 67 (2002).
\bibitem{pseudo}
Cannata F., Ioffe M.V. and Nishnianidze D.N. {\it Phys. Lett. A} {\bf 310} 344 (2003);\\
Cannata F., Ioffe M.V. and Nishnianidze D.N. {\it Theor. Math. Phys.} {\bf 148} 960 (2006)
[Translated from: {\it Teor. Mat. Fiz.} {\bf 148} 102 (2006)]; arXiv hep-th/0512110;\\
Cannata F., Ioffe M.V. and Nishnianidze D.N. {\it Phys. Lett. A} {\bf 369} 9 (2007).
\bibitem{CIN-osc}
Cannata F., Ioffe M.V. and Nishnianidze D.N. {\it J. Math. Phys.} {\bf 51} 022108 (2010).
\bibitem{shape}
Gendenshtein L.E. {\it JETP Lett.} {\bf 38} 356 (1983).
\bibitem{mallow}
Bougie J., Gangopadhyaya A., Mallow J.V., {\it Phys. Rev. Lett.}, {\bf 105} (2010) 210402;\\
Bougie J., Gangopadhyaya A., Mallow J.V., {\it J. Phys. A}, {\bf 44} (2011) 275307.
\bibitem{new}
Cannata F., Ioffe M.V. and Nishnianidze D.N. {J. Phys. A: Math. Gen.} {\bf 35} 1389 (2002).
\bibitem{ioffe1}
Ioffe M.V., {\it J. Phys. A}, {\bf 37} (2004) 10363.
\bibitem{shape-our}
Andrianov A.A., Cannata F., Ioffe M.V. and Nishnianidze D.N. {\it Phys. Lett. A} {\bf 266} 341 (2000).
\bibitem{shape-2011}
Cannata F., Ioffe M.V., Nishnianidze D.N., {\it J. Math. Phys.}, {\bf 52} (2011) 022106.
\bibitem{exact}
Ioffe M.V. and Nishnianidze D.N. {\it Phys. Rev. A} {\bf 76} 052114 (2007).
\bibitem{witten}
Witten E., {\it Nucl. Phys.}, {\bf B188} (1981) 513.
\bibitem{cooper} G. Junker, {\it Supersymmetric Methods in Quantum and Statistical
Physics} (Springer,Berlin,1996);\\
Cooper F., Khare A. and Sukhatme U. {\it Phys. Rep.} {\bf 25} 268 (1995);\\
Bagchi B.K., {\it Supersymmetry in Quantum and Classical Mechanics}, Chapman, Boca Raton, 2001;\\
Fernandez C D.J., {\it AIP Conf. Proc.}, {\bf 1287} (2010) 3.
\bibitem{miller}
Miller W.,Jr., {\it Symmetry and Separation of Variables}, Addison-Wesley Publishing Company, London, 1977;
\bibitem{eisenhart}
Eisenhart L.P., {\it Phys. Rev.}, {\bf 74} (1948) 87.
\bibitem{landau}
Landau L.D. and Lifshits E.M. {\it "Course of Theoretical Physics, Vol 3 (Quantum Mechanics: Non-relativistic Theory)"}
(Elsevier, 1991).
\bibitem{levai}
Levai G., Cannata F. and Ventura A. {\it Phys. Lett. A} {\bf 300} 271 (2002).
\bibitem{prudnikov}
Prudnikov A.P., Brychkov Yu.A. and Marichev O.I. {\it "Integrals and Series. Elementary Functions" Vol.1}
(Gordon and Breach Sci. Publ., New York, 1986).

\end{thebibliography}
\end{document}